\documentclass[aps,reprint,superscriptaddress, twocolumn]{revtex4-1}
\usepackage{blindtext}
\usepackage{graphicx}
\usepackage{upgreek}

\begin{document}
\title{Optical dipole orientation of interlayer excitons in MoSe$_2$-WSe$_2$ heterostacks}
\author{Lukas Sigl}
\email{lukas.sigl@wsi.tum.de}
\author{Mirco Troue}
\affiliation{Walter Schottky Institute and Physics Department, TU Munich, Am Coulombwall 4a, 85748 Garching, Germany}
\author{Manuel Katzer}
\author{Malte Selig}
\affiliation{Institut f\"ur Theoretische Physik, Nichtlineare Optik und Quantenelektronik, Technische Universit\"at Berlin, 10623 Berlin, Germany}
\author{Florian Sigger}
\author{Jonas Kiemle}
\affiliation{Walter Schottky Institute and Physics Department, TU Munich, Am Coulombwall 4a, 85748 Garching, Germany}
\author{Mauro Brotons-Gisbert}
\affiliation{Institute of Photonics and Quantum Sciences, SUPA, Heriot-Watt University, Edinburgh EH14 4AS, UK}
\author{Kenji Watanabe}
\affiliation{Research Center for Functional Materials, National Institute for Materials Science, Tsukuba 305-0044, Japan}
\author{Takashi Taniguchi}
\affiliation{International Center for Materials Nanoarchitectures, National Institute for Materials Science, Tsukuba 305-0044, Japan}
\author{Brian D. Gerardot}
\affiliation{Institute of Photonics and Quantum Sciences, SUPA, Heriot-Watt University, Edinburgh EH14 4AS, UK}
\author{Andreas Knorr}
\affiliation{Institut f\"ur Theoretische Physik, Nichtlineare Optik und Quantenelektronik, Technische Universit\"at Berlin, 10623 Berlin, Germany}
\author{Ursula Wurstbauer}
\affiliation{Institute of Physics, University of M\"unster, Wilhelm-Klemm-Str. 10, 48149 M\"unster, Germany}
\author{Alexander W. Holleitner}
\email{holleitner@wsi.tum.de}
\affiliation{Walter Schottky Institute and Physics Department, TU Munich, Am Coulombwall 4a, 85748 Garching, Germany}

\begin{abstract}
We report on the far-field photoluminescence intensity distribution of interlayer excitons in MoSe$_2$-WSe$_2$ heterostacks as measured by back focal plane imaging in the temperature range between 1.7 K and 20 K. By comparing the data with an analytical model describing the dipolar emission pattern in a dielectric environment, we are able to obtain the relative contributions of the in- and out-of-plane transition dipole moments associated to the interlayer exciton photon emission. We determine the transition dipole moments for all observed interlayer exciton transitions to be $(99\pm 1)\%$ in-plane for R- and H-type stacking, independent of the excitation power and therefore the density of the exciton ensemble in the experimentally examined range. Finally, we discuss the limitations of the presented measurement technique to observe correlation effects in exciton ensembles.
\end{abstract}

\maketitle

\section{INTRODUCTION}
Transition-metal dichalcogenides (TMDs) exhibit a strong light-matter interaction even in the monolayer limit, making them promising candidates for novel two-dimensional (2D) optoelectronic applications \cite{wurstbauerLightMatterInteraction2017}. Due to the weak dielectric screening, the photoluminescence response is dominated by excitons with binding energies on the order of several hundreds of meV \cite{chernikovExcitonBindingEnergy2014}. The fabrication of van der Waals heterostacks with a type II band alignment allows the excitation of so-called interlayer excitons (IXs), where the constituting electron and hole states are located in two adjacent monolayers \cite{geimVanWaalsHeterostructures2013a, novoselov2DMaterialsVan2016a, millerLongLivedDirectIndirect2017, hanbickiDoubleIndirectInterlayer2018, ciarrocchiPolarizationSwitchingElectrical2019, wangGiantValleyZeemanSplitting2020, brotons-gisbertMoireTrappedInterlayerTrions2021}. The reduced wave function overlap in such IXs results in long lifetimes in the order of hundreds of nanoseconds \cite{millerLongLivedDirectIndirect2017}, thus making dense exciton ensembles cooled to lattice temperature experimentally accessible. Such exciton ensembles are ideal to explore the many-body phase diagram in quasi-equilibrium. Due to the bosonic nature of IXs in combination with a permanent out-of-plane electric dipole moment \cite{kiemleControlOrbitalCharacter2020, jaureguiElectricalControlInterlayer2019}, a versatile interaction-driven phase diagram with classical and quantum phases, including superfluidity and a quasi condensation is expected \cite{foglerHightemperatureSuperfluidityIndirect2014, wangEvidenceHightemperatureExciton2019, siglSignaturesDegenerateManybody2020}.

In each monolayer of the heterostacks, the optical selection rules are dominated by a strong spin orbit coupling and a broken inversion symmetry, such that states are typically spin and valley polarized \cite{xiaoCoupledSpinValley2012}. As a result, intralayer excitons with electrons and holes being localized in the same monolayer, can be dipole allowed (bright) with in-plane (IP) or out-of-plane (OP) transition dipole moments, or generally forbidden (dark) depending on the quantum numbers of the contributing single particle states \cite{wozniakExcitonFactorsVan2020, wangColloquiumExcitonsAtomically2018}. In monolayer WS$_2$, WSe$_2$, as well as charge-neutral MoS$_2$, the energetic splitting between bright and dark excitons has been reported to be positive, which implies that the excitonic ground state transition is considered to be momentum or spin forbidden (dark) \cite{kleinControllingExcitonManybody2021, seligExcitonicLinewidthCoherence2016a,wangColloquiumExcitonsAtomically2018,zhangMagneticBrighteningControl2017,robertMeasurementSpinforbiddenDark2020,molasBrighteningDarkExcitons2017,luMagneticFieldMixing2019,feierabendBrighteningSpinMomentumdark2020,robertMeasurementSpinforbiddenDark2020}. In contrast, the  energetically lowest exciton transition in MoSe$_2$ is reported to be bright with an in-plane (IP) optical dipole moment \cite{seligUltrafastDynamicsMonolayer2019, wangInPlanePropagationLight2017,luMagneticFieldMixing2019, brotons-gisbertOutofplaneOrientationLuminescent2019}.
For heterostacks, the combination of band hybridization \cite{kiemleControlOrbitalCharacter2020}, strong spin-orbit interaction, atomic reconstruction of the interface \cite{rosenbergerTwistAngleDependentAtomic2020}, and the possible formation of Moir\'{e} superlattices \cite{seylerSignaturesMoiretrappedValley2019} makes an accurate description of the exciton transitions more demanding. First theoretical work presents calculated band structures and optical selection rules particularly for MoSe$_2$-WSe$_2$ heterostacks \cite{wozniakExcitonFactorsVan2020,yuBrightenedSpintripletInterlayer2018,gillenInterlayerExcitonicSpectra2021a}. For the corresponding IXs, a finite oscillator strength is predicted for both IP ($\sigma\pm$) and OP (z) transition dipoles \cite{yuBrightenedSpintripletInterlayer2018}. However, so far no signatures of $z$-polarized IX transitions have been reported and the quantitative contribution of OP transition dipoles to the overall photoluminescence remains unclear.

\begin{figure*}[t]
    \centering
    \includegraphics[scale=0.8]{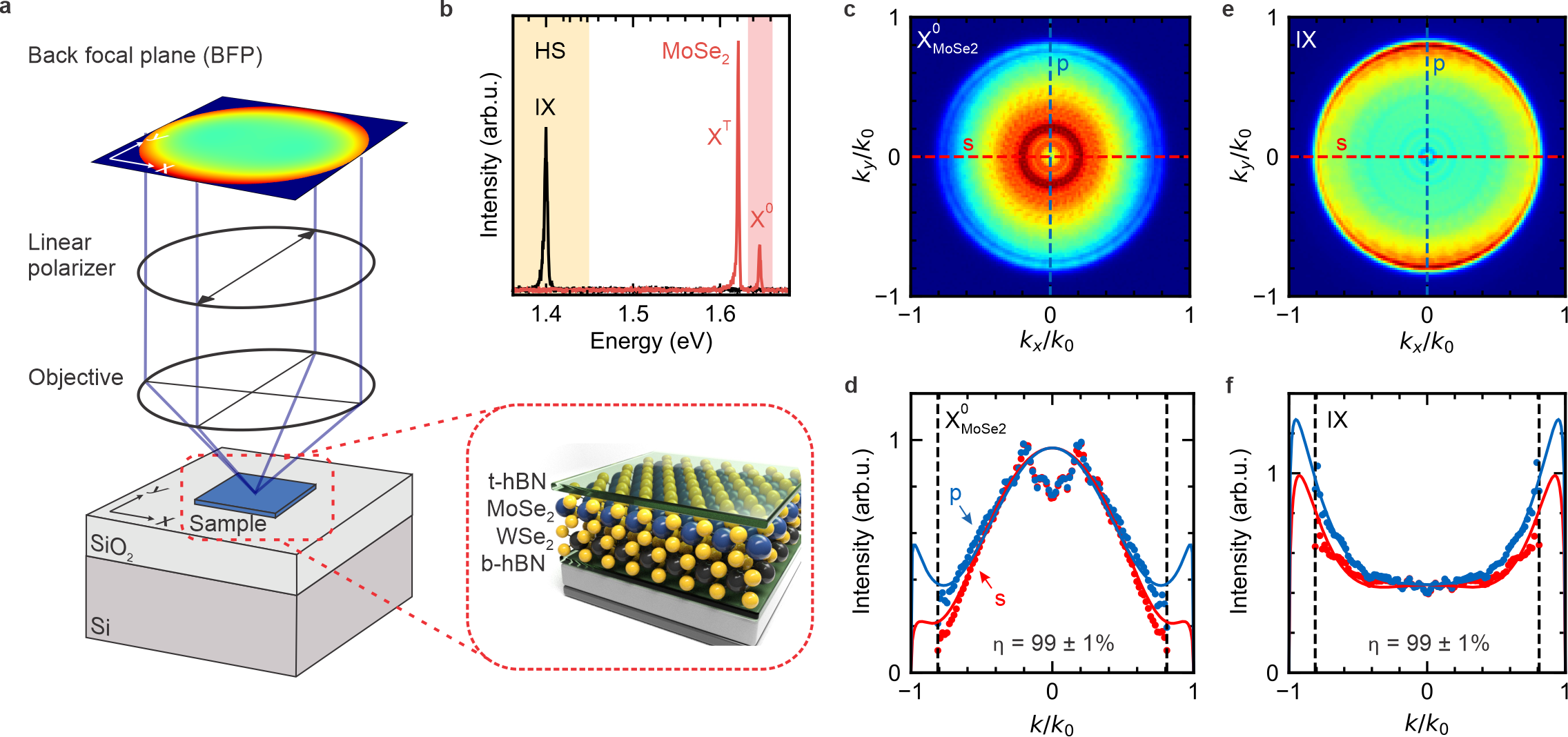}
    \caption{\textbf{a} Sketch of the back focal plane (BFP) imaging principle on 2D materials. The inset on the lower right displays a scheme of the sample structure. A MoSe$_2$-WSe$_2$ hetreobilayer is encapsulated within a top- and bottom hBN layer. The substrate is silicon (Si) with a 290 nm thick SiO$_2$ buffer layer. \textbf{b} Comparison of photoluminescence signal on the heterostack (black) and monolayer MoSe$_2$ (red). \textbf{c} Intensity normalized $k$-space emission pattern of intralayer excitons X$^0$ of monolayer MoSe$_2$. The photoluminescence is filtered by an optical bandpass filter (red area in \textbf{b}) and analysed by a linear polarizer. The in-plane photon wave vector is normalized to $k_0$ in air. \textbf{d} Cross section of the emission pattern in \textbf{c} along the directions parallel (blue) and perpendicular (red) to the linear polarizer. The blue and red solid lines correspond to a model fit revealing a $(99\pm 1)\%$ IP optical dipole moment. The deviation of the model from the measurements around $k$ = 0 is caused by impurities in the optical path. \textbf{e} Intensity normalized $k$-space emission pattern of interlayer excitons (IXs). \textbf{f} Cross section of the emission pattern in \textbf{e} along the directions parallel (blue) and perpendicular (red) to the linear polarizer. Fitting the source term model (blue and red solid lines), we obtain an optical dipole orientation of $\alpha=87.3^\circ$, which corresponds to $\eta = (99\pm 1)\%$.}
    \label{fig:1}
\end{figure*}

In this work, we experimentally determine the optical dipole orientation of interlayer excitons (IXs) in R- and H-type MoSe$_2$-WSe$_2$ heterostacks via back focal plane (BFP) imaging at low temperatures. So far, room temperature experiments resolved the optical transition dipole moment from the photoluminescence far-field distribution of single molecules \cite{liebSinglemoleculeOrientationsDetermined2004} and excitons in TMD monolayers \cite{schullerOrientationLuminescentExcitons2013,brotons-gisbertOutofplaneOrientationLuminescent2019,zhangMagneticBrighteningControl2017,wangInPlanePropagationLight2017,schneiderDirectMeasurementRadiative2020}. At low temperatures, single photon emitters and excitons in WSe$_2$ have been characterized in this manner \cite{luoExcitonDipoleOrientation2020a}. For the heterostacks, we quantify the contribution of IP and OP optical dipole moments to the IX photoluminescence by utilizing an analytical model describing a dipole emission pattern in a dielectric environment in dependence of temperature and laser excitation power.

\begin{figure*}[t]
    \centering
    \includegraphics[scale=0.8]{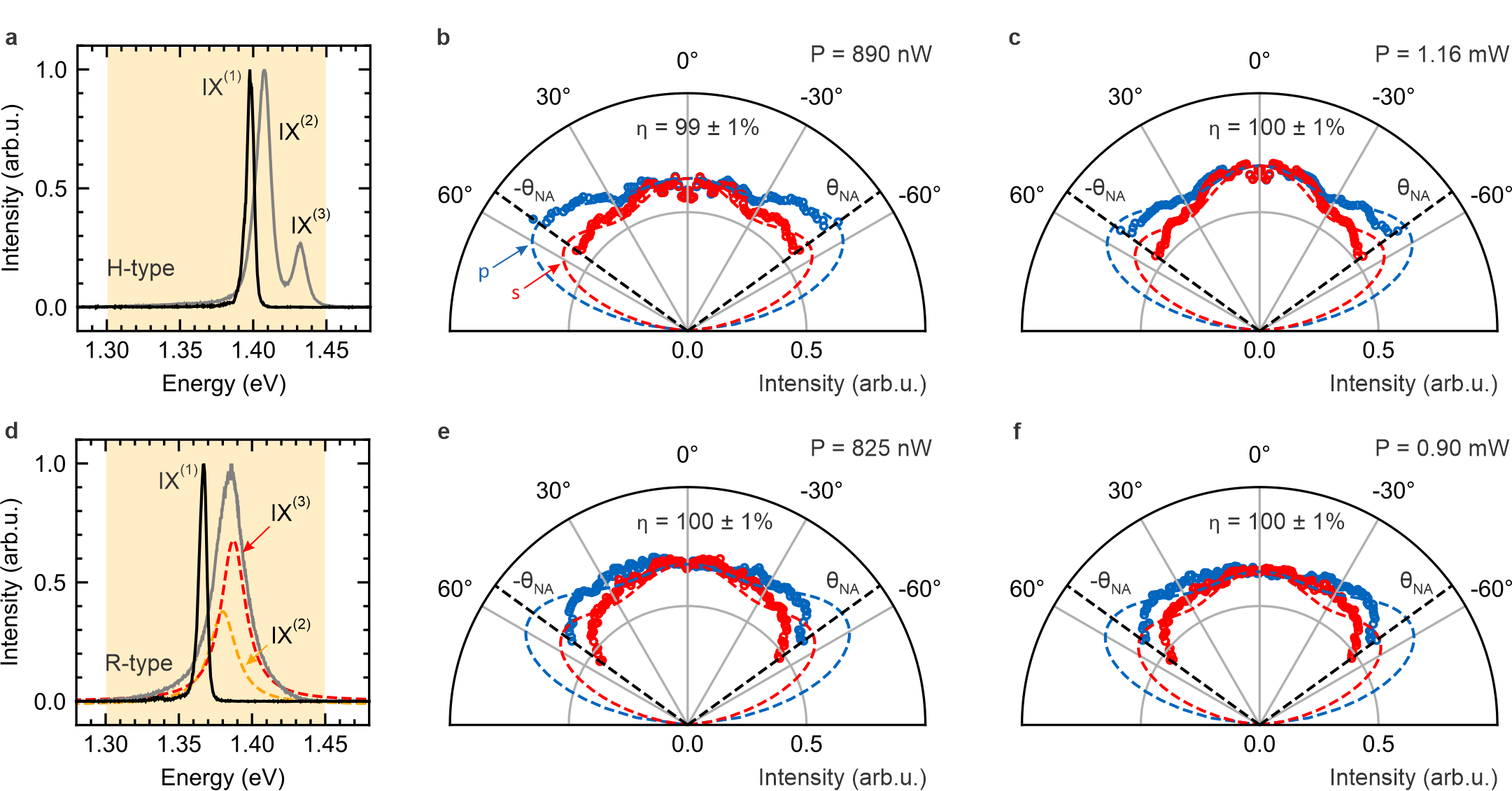}
    \caption{\textbf{a} Low temperature (T=10 K) photoluminescence spectra of interlayer excitons (IX$^{(1)}$,IX$^{(2)}$,IX$^{(3)}$) on a heterostack with H-type stacking ($\phi\simeq 60^\circ$) at two different excitation powers, P = 890 nW (black) and P = 1.16 mW (grey). Optical filters are used to suppress photoluminescence signal outside the wavelength range of $850-900$ nm, indicated by the yellow shaded area. The corresponding s- and p-polarized far-field emission pattern of the spectra in \textbf{a} as a function of the emission angle $\theta^{\text{out}}$ are presented in \textbf{b} and \textbf{c}, respectively. The black dashed lines indicate the maximum collection angle ($\theta_{\text{NA}}$) provided by the NA of the objective. Red and blue dashed lines represent the resulting fits by the source term model. The relative contribution of IP transition dipole moments is denoted by $\eta$. \textbf{d} Photoluminescence spectra ($T = 1.7$ K) of interlayer excitons on a heterostack with R-type stacking ($\phi\simeq 0^\circ$) at excitation powers P = 825 nW (black) and P = 0.90 mW (grey). The spectrum at higher excitation power is fitted by two Lorentzians, referred to as IX$^{(2)}$ (orange) and IX$^{(3)}$ (red). \textbf{e,f} The corresponding far-field emission pattern of the spectra in \textbf{d} as a function of the emission angle $\theta^{\text{out}}$.}
    \label{fig:2}
\end{figure*}
\section{RESULTS}

In our experiments, the MoSe$_2$-WSe$_2$ heterostacks are cooled down to a lattice temperature between 1.7 K and 20 K in a closed cycle helium cryostat. We optically excite the samples at $\lambda = 639$ nm (1.95 eV). A low-temperature microscope objective (NA = 0.81) focuses the light to a diffraction-limited spot and the emitted light is collected by the same objective. Two achromatic lenses inside the cryostat act as relay lenses. We use an achromatic tube lens outside of the cryostat to map a real-space image of our sample onto a charge-coupled device. By inserting a Bertrand lens in front of the charge-coupled device, we can image the back focal plane (BFP) [Fig. 1(a)].
The angular emission pattern depends on the orientation of the optical transition dipole moment and is encoded in the resulting intensity distribution. We can distinguish between s- and p-polarized emission by inserting a linear polarizer after the objective. Finally, we compare the experimental data to an analytical model, as described later in the text to extract the dipole orientation, which can be depicted as a linear combination of IP and OP optical dipole moments \cite{brotons-gisbertEngineeringLightEmission2018}.

Fig. 1(b) shows an exemplary photoluminescence spectrum of hBN encapsulated monolayer MoSe$_2$ at a lattice temperature of $T=1.7 $ K (red). We observe bright intralayer exciton emission attributed to the charge-neutral 1s exciton X$^0$ at 1.645 eV and the charged trion X$^{\text{T}}$ at 1.621 eV, in very good agreement to current reports in literature \cite{ciarrocchiPolarizationSwitchingElectrical2019}. In contrast, on the heterostack region of this first sample (I), intralayer exciton emission is quenched most likely due to an effective charge transfer between the layers \cite{millerLongLivedDirectIndirect2017, kiemleControlOrbitalCharacter2020}. Instead, we observe an emerging emission line around 1.4 eV (black), which is attributed to the formation of IXs with the electron residing in MoSe$_2$ and the hole in WSe$_2$ \cite{siglSignaturesDegenerateManybody2020}.

We start showing the experimental $k$-space emission profiles of the neutral 1s intralayer exciton of monolayer MoSe$_2$ in Fig. 1(c). The photoluminescence is filtered by an optical bandpass filter [marked by the red area in Fig. 1(b)] to suppress other luminescence, e.g. from X$^{\text{T}}$. To reduce noise and to average out anisotropies along the optical path, we record 36 BFP images by rotating the polarizer in steps of $10^\circ$ over $360^\circ$. The final image is then obtained by rotating and averaging all images \cite{brotons-gisbertOutofplaneOrientationLuminescent2019}. The vertical ($k_y/k_0$) and horizontal ($k_x/k_0$) axes in Fig. 1(c) represent the orthogonal components of the in-plane photon wave vector ($k_0\cdot \sin(\theta^{\text{out}}))$, with the emission angle $\theta^{\text{out}}$ and $k_0 = E/\hbar c$ the photon wave vector in air. Photons with an in-plane wave vector larger than the numerical aperture of the objective are not collected (dark blue region).

We simulate the $k$-dependent dipole emission profile of excitons in our heterostacks by employing the model proposed by Benisty et al. \cite{benistyMethodSourceTerms1998}. The model combines a transfer matrix method with dipole emission source terms originating from the source layer. We assume an isotropic radial distribution of incoherent dipoles that exhibit an angle $\alpha$ with respect to the out-of-plane direction ($z$). The s- and p-polarized components of the outside electric field $E_{\text{p,s}}^{\text{out}}( \alpha, \theta^{\text{out}})$ are calculated separately for each emission angle $\theta^{\text{out}}$. Finally, the $k$-space intensity profiles for directions parallel ($I_{\text{p}}$) and perpendicular ($I_{\text{s}}$) to the collection polarizer are calculated as
\begin{equation}
    I_{\text{p,s}}(\alpha, \theta^{\text{out}}) = \frac{1}{\cos(\theta^{\text{out}})}\left|E_{\text{p,s}}^{\text{out}} (\alpha, \theta^{\text{out}}) \right|^2,
\end{equation}
with the standard apodization factor $\cos^{-1}(\theta^{\text{out}})$ to consider the energy conservation after collimating the light with an objective \cite{liebSinglemoleculeOrientationsDetermined2004}. The contribution of IP dipole moments to the total emission intensity $I$ is determined by the following expression
\begin{equation}
    \eta := \frac{I_{\text{IP}}}{I} = \sin^2(\alpha),
    \label{eq:ip}
\end{equation}
where $\alpha$ represents the dipole orientation. We assume a confidence level of $\pm 1\%$ for the extracted $\eta$ \cite{brotons-gisbertOutofplaneOrientationLuminescent2019}. Since we independently determine the thickness of each layer as well as the refractive indices \cite{greenSelfconsistentOpticalParameters2008, leeRefractiveIndexDispersion2019, rodriguez-demarcosSelfconsistentOpticalConstants2016}, we can use the source term model to fit the observed BFP images, with $\alpha$ as the only free parameter. From the optimized $\alpha$, we obtain the underlying contributions from IP/OP dipole moments to the total emission according to Eq. (\ref{eq:ip}).

Fig. 1(d) presents the vertical and horizontal cross-sections of the normalized $k$-space emission pattern from Fig 1(c). They correspond to the directions parallel and perpendicular to the linear polarizer. The reduced intensity in the range $-0.2<k/k_0 <0.2$ is caused by an impurity in the optical path, most probably due to a small dust particle/dirt on an optical lens. The quenched signal by the impurity is not affected by the rotating polarizer and thus can be easily identified. Consequently, we exclude this range in the model fit. The result is presented by the solid red and blue lines and corresponds to a $(99\pm 1)\%$ IP dipole moment [$\alpha=85^\circ$, cf. Eq. (\ref{eq:ip})], as expected for the neutral intralayer exciton $\text{X}^0$ in monolayer MoSe$_2$ \cite{wozniakExcitonFactorsVan2020, yuBrightenedSpintripletInterlayer2018, brotons-gisbertOutofplaneOrientationLuminescent2019}. 

The $k$-space emission pattern for IXs of sample I is displayed in Fig. 1(e), with the corresponding cross-sections in Fig. 1(f). In contrast to the results on the intralayer excitons, we observe enhanced intensities at larger $k$-vectors compared to $k=0$. However, from the model fit indicated as blue and red solid lines, we again obtain a $(99 \pm 1)\%$ IP dipole moment [$\alpha=87^\circ$, cf. Eq. (\ref{eq:ip})]. The apparently differing BFP images for intra- and interlayer excitons solely result from the differing emission wavelengths of intra- and interlayer excitons, since wavelength-dependent refractive indices and interference effects in the multilayer structure alter the far-field intensity depending on the emission angle \cite{brotons-gisbertEngineeringLightEmission2018}.

In the next step, we characterize the emission profile of the IXs as a function of density and stacking angle. Fig. 2(a) shows the photoluminescence spectra of sample (I) at excitation powers of $P=890$ nW in black and $P=1.16$ mW in grey. The stacking type is identified by measuring the effective Lande g-factor ($g = -14.8\pm 0.2$), which we attribute to H-type stacking \cite{seylerSignaturesMoiretrappedValley2019}. From second-harmonic generation of each layer, we determine the rotational alignment $\Delta\phi$ to be $(0 \pm 1)^\circ$. We observe three different IX emission lines, labeled as IX$^{(1)}$, IX$^{(2)}$ and IX$^{(3)}$ from lowest to highest energy. At low excitation powers, only IX$^{(1)}$ is observed in the spectrum. For increasing excitation powers, IX$^{(2)}$ and IX$^{(3)}$ emerge and finally dominate the photoluminescence, as reported by us before in ref. \cite{siglSignaturesDegenerateManybody2020}.

Figs. 2(b) and (c) show the corresponding far-field emission pattern as a function of emission angle $\theta^{\text{out}}$. The maximum detection angle $\pm\theta_{\text{NA}}$ is determined by the objective's NA and constitutes to $\pm\theta_{\text{NA}}= 54.1^\circ$ (NA = 0.81). Independent of the excitation power, the fits to the observed $k$-space images (dashed lines) reveal $\eta=(99\pm1)\%$ within the given experimental uncertainties. The slight deviations between low [Fig. 2(b)] and high excitation powers [Fig. 2(c)] are captured in the model by shifts in the IX emission wavelengths. We note that a second sample (II) with H-type stacking and similar IX emission shows the same robust IP contribution $\eta$. 

In Fig. 2(d), we present the photoluminescence characteristics of a third sample (III) with R-type stacking ($g  =+4.8\pm 0.1, \Delta\phi=(0\pm 1)^\circ$), with excitation powers of $P=825$ nW (black) and $P=0.96$ mW (grey). Similar to the previous sample, we observe three emission lines with the same power-dependent relative intensities. The model fit reveals $\eta = 100\%$ for all excitation powers; compare Figs. 2(e) and (f).

In Fig. 3(a), we present the normalized photoluminescence spectra of sample I for increasing temperatures from $1.7$ K up to $20$ K. The excitation power is kept constant at $P=5.6~\upmu$W. Overall, the emission of IX$^{(1)}$ dominates the photoluminescence in the chosen experimental parameter space. However, for increasing temperature, the intensity of IX$^{(1)}$ exponentially decreases \cite{siglSignaturesDegenerateManybody2020}, and thus, the relative intensities of emission lines IX$^{(2)}$ and IX$^{(3)}$ increase. Nevertheless, the simultaneously revealed IP dipole moment contribution of $\eta = 98-100\%$ stays constant within the uncertainties, as depicted in Fig. 3(b).

\begin{figure}[t]
    \includegraphics[scale=0.8]{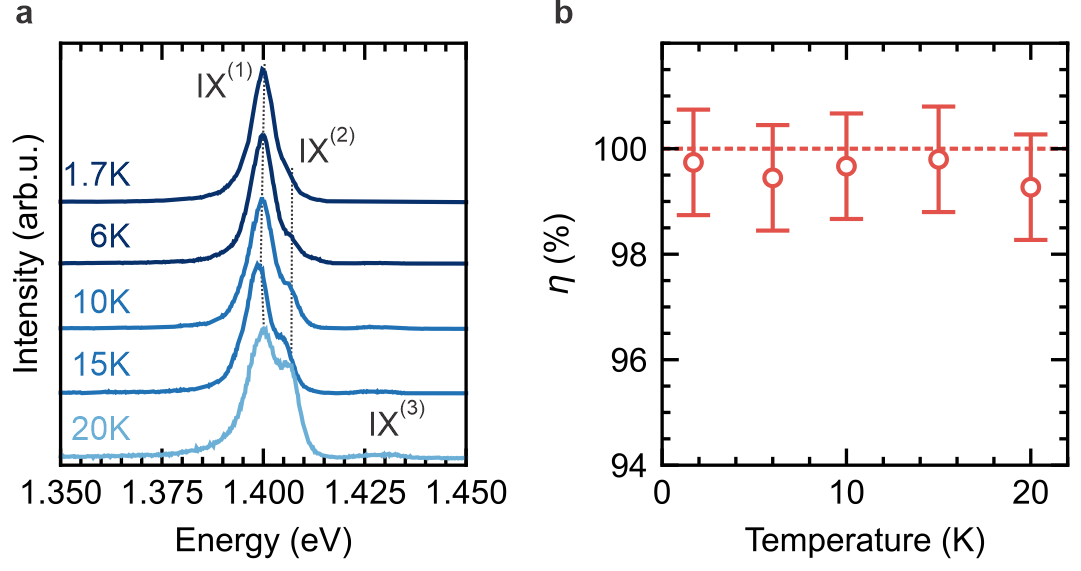}
    \caption{\textbf{a} Normalized photoluminescence spectra of IXs from 1.7 K to 20 K. Excitation power is set to $P=5.6$ $\upmu$W. Optical filters are used to suppress photoluminescence signal outside the wavelength range of $850-900$nm. Overall, emission of IX$^{(1)}$ dominates the spectra for the chosen laser intensity. For higher temperatures, the relative intensity of emission line IX$^{(2)}$ increases. \textbf{b} Corresponding relative contribution of IP optical transition dipole moments $\eta$ as a result of the source term model fit.}
    \label{fig:3}
\end{figure}

\section{DISCUSSION}
On all our samples we observe three different IX emission lines. Several suggestions have been made in literature to address the origin of the different transitions, considering momentum direct \cite{ciarrocchiPolarizationSwitchingElectrical2019, wangGiantValleyZeemanSplitting2020, brotons-gisbertMoireTrappedInterlayerTrions2021} and indirect transitions \cite{hanbickiDoubleIndirectInterlayer2018}. Momentum-indirect exciton states require additional momentum during the radiative recombination process, which may be provided by phonons \cite{bremPhononAssistedPhotoluminescenceIndirect2020}. The phonon-assisted radiative recombination couples to the light field via virtual transitions at the K valley. Consequently, the corresponding emission line in the spectrum inherits the polarization of the K point. In turn, we do not expect different polarizations of momentum direct and indirect transitions, which makes them indistinguishable in the BFP images.

Our results show that the observed BFP emission patterns of IXs in MoSe$_2$-WSe$_2$ hetero-bilayers can be consistently simulated by an optical dipole approximation, where the dipole radiation is modified by a layered dielectric environment \cite{benistyMethodSourceTerms1998}. From the model, we determine that the photoluminescence of all observed IX$^{(1)}$, IX$^{(2)}$, and IX$^{(3)}$ is dominated by IP optical dipole moments. We obtain an upper limit for the OP contribution to be $2\%$ of the total oscillator strength, independent of the stacking type (H- or R-type) and of the here studied excitation powers and temperatures. 

The findings are consistent with previous experiments reporting on a circularly polarized photoluminescence from IXs \cite{hanbickiDoubleIndirectInterlayer2018, ciarrocchiPolarizationSwitchingElectrical2019}. From theory, IP and OP optical transition dipoles are predicted for both R- and H-type stacking \cite{yuBrightenedSpintripletInterlayer2018, wozniakExcitonFactorsVan2020}. We note that structural deformations are likely to occur in our samples due to an atomic reconstruction \cite{linLargeScaleMappingMoire2021}, since all measured samples exhibit a rotational alignment of $\Delta\phi \simeq 0^\circ$ within the given uncertainties. As a result, the ideal Moir\'{e} pattern is reconstructed and the crystal structure is expected to exhibit laterally extended areas of high symmetry \cite{rosenbergerTwistAngleDependentAtomic2020}. Exemplary, for $\phi =60^\circ$ (H-type), the stacking configuration H$_h^h$ is expected to cover most of the heterostack area \cite{wozniakExcitonFactorsVan2020}. In turn, the oscillator strength related to OP transition dipole moments is expected to be negligible \cite{yuBrightenedSpintripletInterlayer2018}, as is consistent with our experiments.

In a previous work, we attributed IX$^{(1)}$ as in sample I to a possible excitonic many-body state which portrays itself in several criticalities \cite{siglSignaturesDegenerateManybody2020}, particularly as a function of temperature and exciton density. As can be seen in Figs. 2(b) and (c), the measured exciton distributions do not show a pronounced occupation of the energetically lowest states around $k=0$, even at the lowest measured temperatures of 1.7 K. For comparison, in a similar temperature range, sharp distributions below a momentum of $10^{-3} ~\text{nm}^{-1}$  were reported for exciton-polariton condensates realized in a CdTe/CdMgTe microcavity grown by molecular beam epitaxy \cite{kasprzakBoseEinsteinCondensation2006}. For polaritons, however, the condensation takes place in the energetically lower polariton branch of a microcavity, where the polariton mass is only on the order of 10$^{-4}$ free electron masses. The small effective mass results in an extremely narrow polariton distribution in momentum space e.g. with a FWHM in the range of $3.5 \cdot 10^{-4} ~\text{nm}^{-1}$ at $\mu/k_BT=-0.05$. For the samples studied here, the IXs have masses on the order of the free electron mass ($M \approx 1.04~m_e$ in MoSe$_2$-WSe$_2$ heterostacks) \cite{kormanyosTheoryTwodimensionalTransition2015}. In turn, the actual exciton distribution is much broader, with an expected FWHM in the range of 0.035 nm$^{-1}$ for the same discussed condensation conditions, i.e. $\mu/k_BT = -0.05$. Additionally, the observation of the involved exciton distributions is limited by the photon momentum $k_0 \simeq 0.007$ nm$^{-1}$ (at an emission energy of ~1.4 eV). Consequently, the exciton distribution can be expected to be flat in the radiative window even for high densities, as we observe in Figs. 1(e) and (f) but in contrast to the polariton case. The above considerations apply to both IXs being direct or indirect in $k$-space~\cite{millerLongLivedDirectIndirect2017}. For IXs that are indirect in momentum space, the radiative decay has to be assisted by additional phonon scattering events or disorder \cite{bremPhononAssistedPhotoluminescenceIndirect2020}, both of which lead to an extra broadening of the observed spectra beyond the actual exciton distribution. While the dispersion and coupling strength of intervalley transitions do not significantly depend on the momentum of the involved phonon in the range of the radiative light cone \cite{jinIntrinsicTransportProperties2014}, the phonon scattering events smear out the actually narrow exciton dispersion during the phonon-assisted emission process leading to a relatively flat profile in the emission. As a consequence, we do not expect different BFP emission patterns between a weakly interacting exciton gas and a correlated exciton ensemble under the given experimental conditions. Moreover, we note that interference effects in the heterostacks at specific wavelengths can further overlay the emission pattern [cf. discussion of Figs. 1(d) and (f)], and as far as the model is concerned, it only considers an incoherent optical dipole with a single frequency to describe the exciton photoluminescence \cite{benistyMethodSourceTerms1998}. Last but not least, we give an upper boundary for the IX densities as explored in this report for samples I, II, and III. The maximum density for IX$^{(1)}$ can be estimated to be in the order of 3~$\cdot~10^{11}$~cm$^{-2}$ before IX$^{(2)}$ and IX$^{(3)}$ appear in the spectrum (cf. \cite{siglSignaturesDegenerateManybody2020}). Future experiments at a possibly lower exciton temperature and similar or higher densities will show whether BFP imaging is able to resolve the IX distributions at very small $k$ indicating possible many-body phases of IX ensembles.

\section{CONCLUSION}
In summary, we experimentally investigated the far-field photoluminescence intensity distribution of interlayer excitons (IXs) in three different MoSe$_2$-WSe$_2$ heterostacks based on back focal plane imaging. A model assuming a classical dipole radiation and being modified by a layered dielectric environment successfully reproduces the observed patterns. From the model, we determine that the photoluminescence of all observed IXs is dominanted by IP optical dipole moments. We obtain an upper limit for the OP contribution to be $2\%$ of the total oscillator strength, independent of the stacking type (H- or R-type) and the here studied excitation powers and temperatures. 

\section{ACKNOWLEDGMENTS}
We gratefully acknowledge financial support by the Deutsche Forschungsgemeinschaft (DFG) via Projects No. 
HO 3324 / 9-2 and WU  637 / 4-2 and the excellence clusters Munich Center for Quantum Science and Technology (MCQST) - EXC-2111390814868 and e-conversion - EXC 2089/1-390776260. K.W. and T.T. acknowledge support from the Elements Strategy Initiative conducted by the MEXT, Japan (Grant Number JPMXP0112101001) and JSPS KAKENHI (Grant Numbers 19H05790 and JP20H00354). M.K., M.S. and A.K. gratefully acknowledge funding by the Deutsche Forschungsgemeinschaft (DFG) - project number 182087777 - SFB 951.

\section{APPENDIX}

\subsection{Sample characterization}
\begin{table}[h]
    \centering
    \begin{tabular}{|c|c|c|c|c|}
    \hline
         Sample & SiO$_2$ (nm) & b-hBN (nm) & t-hBN (nm) \\
         \hline
          I    &  $295.6\pm 0.9$ &     $8.7\pm0.4$ & $13.4\pm0.4$\\
          II    &  $294.5\pm 0.1$ &     $15.6\pm0.2$ & $13.9\pm0.1$  \\
          III    &  $286.6\pm 0.4$ &     $18.5\pm0.1$ & $16.0\pm0.1$ \\
          \hline
    \end{tabular}
    \caption{List of measured layer thicknesses for SiO$_2$, bottom- and top-hBN for all samples.}
    \label{tab:tab_1}
\end{table}
Table \ref{tab:tab_1} gives an overview of the layer thicknesses for the presented samples. The parameters are required to adequately model the dipole emission pattern in the source term model. We determine the thicknesses of top- an bottom hBN by atomic force microscopy (AFM) and the SiO$_2$ layer by whitelight reflectometry. The wavelength dependent refractive indices are taken from references ~\cite{greenSelfconsistentOpticalParameters2008, leeRefractiveIndexDispersion2019, rodriguez-demarcosSelfconsistentOpticalConstants2016}.



\newpage

\bibliography{BFP_interlayer}

\end{document}